\documentclass{aastex631}

\usepackage{amsmath}
\usepackage{graphicx}

\begin{document}

\title{A Data-Driven Approach for Mitigating Dark Current Noise and Bad Pixels in Complementary Metal Oxide Semiconductor Cameras for Space-based Telescopes}

\correspondingauthor{Peng Jia, Yushan Li}
\email{robinmartin20@gmail.com}
\email{liys9378@gmail.com}

\author[0000-0001-6623-0931]{Peng Jia}
\affiliation{College of Electronic Information and Optical Engineering, Taiyuan University of Technology, Taiyuan, 030024, China}

\author{Chao Lv}
\affiliation{College of Electronic Information and Optical Engineering, Taiyuan University of Technology, Taiyuan, 030024, China}

\author{Yushan Li}
\affiliation{College of Electronic Information and Optical Engineering, Taiyuan University of Technology, Taiyuan, 030024, China}
\affiliation{Department of Physics, The University of Hong Kong, Hongkong, 999077, China}

\author{Yongyang Sun}
\affiliation{College of Electronic Information and Optical Engineering, Taiyuan University of Technology, Taiyuan, 030024, China}

\author{Shu Niu}
\affiliation{Origin Space Co., Ltd., Nanjing, 210094, China}

\author{Zhuoxiao Wang}
\affiliation{Origin Space Co., Ltd., Nanjing, 210094, China}



\begin{abstract}
In recent years, there has been a gradual increase in the performance of Complementary Metal Oxide Semiconductor (CMOS) cameras. These cameras have gained popularity as a viable alternative to charge-coupled device (CCD) cameras in a wide range of applications. One particular application is the CMOS camera installed in small space telescopes. However, the limited power and spatial resources available on satellites present challenges in maintaining ideal observation conditions, including temperature and radiation environment. Consequently, images captured by CMOS cameras are susceptible to issues such as dark current noise and defective pixels. In this paper, we introduce a data-driven framework for mitigating dark current noise and bad pixels for CMOS cameras. Our approach involves two key steps: pixel clustering and function fitting. During pixel clustering step, we identify and group pixels exhibiting similar dark current noise properties. Subsequently, in the function fitting step, we formulate functions that capture the relationship between dark current and temperature, as dictated by the Arrhenius law. Our framework leverages ground-based test data to establish distinct temperature-dark current relations for pixels within different clusters. The cluster results could then be utilized to estimate the dark current noise level and detect bad pixels from real observational data. To assess the effectiveness of our approach, we have conducted tests using real observation data obtained from the Yangwang-1 satellite, equipped with a near-ultraviolet telescope and an optical telescope. The results show a considerable improvement in the detection efficiency of space-based telescopes.
\end{abstract}



\section{Introduction} \label{sec:intro}
Detectors play a crucial role in modern astronomical observations. Currently, there are two primary types of detectors employed for ultraviolet, optical, and infrared astronomical observations: charge-coupled devices (CCD) and complementary metal oxide semiconductor (CMOS) detectors. CCD detectors have been used in astronomical observations for several decades \citep{mackay1986charge}, due to their large detection area, low noise levels, and uniform photon-electron response \citep{fossum2014review}. However, CCD cameras do have certain drawbacks, including slow read-out speeds and high costs. On the other hand, CMOS detectors, commonly found in commercial cameras, present a viable alternative. However, their application in astronomical observations is limited by a specific issue. Each pixel in a CMOS camera contains its own amplifier and A/D conversion circuit \citep{howell2006handbook}. Consequently, the dark current and quantum efficiency exhibit nonuniformity between pixels, introducing additional noise that can adversely affect the detection efficiency, photometry, and astrometry accuracy \citep{mahato2018measuring}.\\

Due to recent advancements in semiconductor technology, CMOS cameras have been able to achieve increasingly impressive performance. Certain properties of CMOS cameras, such as readout noise, quantum efficiency, and dark current levels, now rival or surpass those of CCD cameras. In addition, CMOS cameras offer advantages such as Faster readout speeds, lower costs, and reduced power consumption \citep{litwiller2001ccd}, making them the preferred choice for numerous astronomical applications. Examples include their utilization in lucky imaging systems \citep{law2006lucky} and as detectors in wavefront sensors \citep{basden2015analysis}. The installation of CMOS cameras in space-based wide-field telescopes is also being considered \citep{karpov2021characterization}. These telescopes are typically employed for time-domain astronomy that involves the detection of exoplanets, supernovae, or near-Earth objects. However, the limited space and energy resources available on satellite platforms pose challenges for CMOS cameras in maintaining optimal operating conditions. This gives rise to two issues:\\
1. Temperature fluctuations can cause variations in the dark current of CMOS cameras. Due to the independent read-out and A/D conversion circuits in each pixel, the relationship between temperature and dark current can differ between pixels.\\
2.Cosmic radiation can damage pixels in CMOS cameras, leading to the generation of bad pixels that exhibit noticeable deviations from their expected performance.\\

Dark current and bad pixels can introduce false detection results and reduce the accuracy of photometry and astrometry. Various methods have been proposed in industries to measure dark current \citep{king1997dark, cassiago2000stability, guarnieri2022smu, liu2022hybrid} or model the imaging process of CMOS cameras using physical models \citep{konnik2014high}. However, these dark current measurement methods require additional circuits, which are not practical to implement after the camera has been assembled for satellites. Additionally, CMOS camera simulators generate simulated images using a forward model, which cannot replicate the dark current specific to a particular CMOS camera based on telemetry data. Traditionally, dark current data can be obtained during real observations, and post-processing methods can be employed to mitigate the effects caused by dark current or bad pixels \citep{rauscher2011dark}. However, this approach requires additional observations on orbit to acquire calibration data, resulting in significant data transfer and observation resource costs.\\

In this paper, we present a data-driven approach for the CMOS camera to reduce effects caused by dark current and bad pixels. Our approach includes pixel clustering step and a dark current - temperature fitting step based on physical model proposed in \citet{lv2022bad}. In terms of the physical aspect, it is expected that the relationship between dark current levels and temperature should exhibit a consistent pattern for all pixels. Additionally, the relationship between dark current levels and temperature for the same pixel in various observations should also adhere to the same trend. Therefore, we can use data obtained from laboratory tests to build relations between dark current and temperature. To achieve this, we initially acquire ground-based test data that simulates space observation conditions. We then categorize these pixels based on the relation between temperature and dark current level. Finally, we employ the Arrhenius law to establish the relationship between dark current levels and temperature \citep{widenhorn2002temperature}. In real applications, our approach serves as a tool for processing real observational data, enabling the estimation of dark current levels and the identification of bad pixels. Details regarding the telescope and camera employed in this study will be provided in Section~\ref{sec:ins}. In Section~\ref{sec:CMOS}, we will introduce the concept of our approach. Practical applications will be presented in Section~\ref{sec:realapp}. Finally, in Section~\ref{sec:con}, we will present our conclusions and outline our future work.\\

\section{Instruments and Data} \label{sec:ins}
\begin{figure} [ht]
\label{fig:through}
   \begin{center}
   \begin{tabular}{c} 
   \includegraphics[height=5cm]{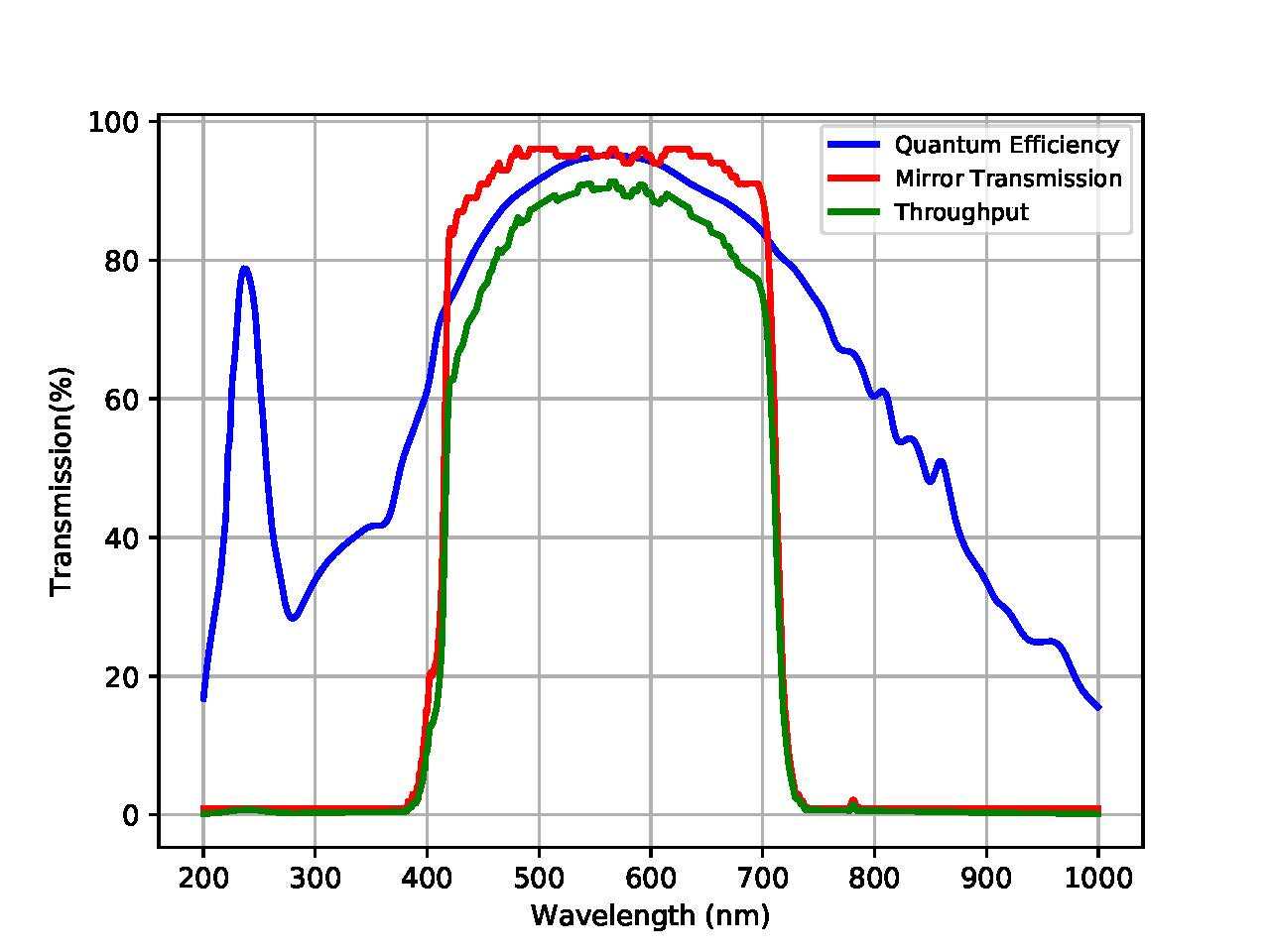}
   \includegraphics[height=5cm]{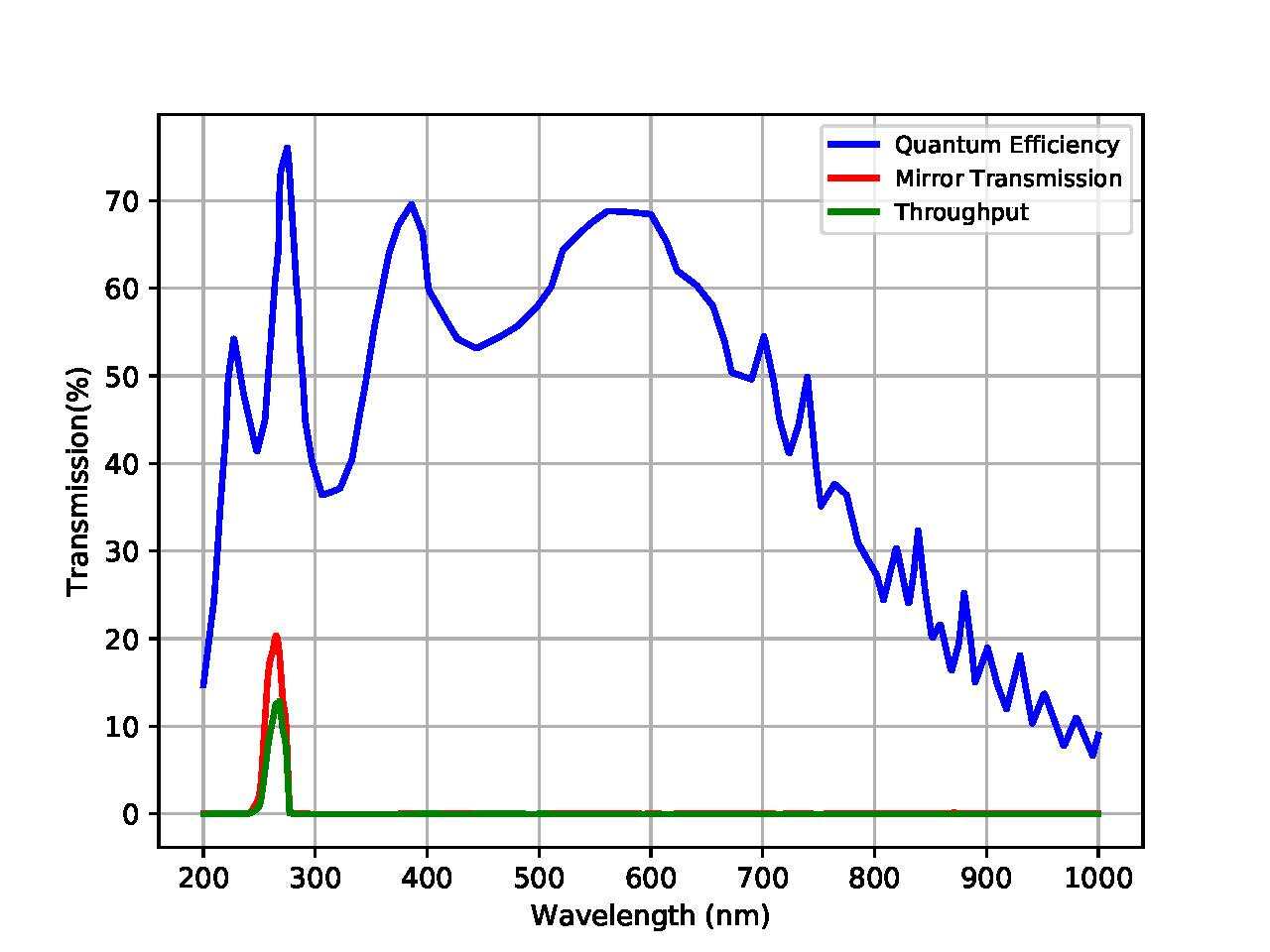}
   \end{tabular}
   \end{center}
    \caption
   { \label{fig:00}The left panel shows the throughput of the optical telescope and the right panel shows that of the near-ultraviolet telescope.}
 \end{figure}

In this paper, we utilize our method to process data acquired from Yangwang-1 (IAU Space Telescope No. C59).The Yangwang-1 is a 4U satellite maintained by Origin Space Co., Ltd. It was launched on June 11th, 2021. The satellite houses two telescopes: an optical telescope and a near--ultraviolet telescope. The key parameters of these two telescopes are provided in Table~\ref{tab:parameter} and the throughput of these telescopes is shown in Figure~\ref{fig:00}. As indicated in the table, both telescopes boast a large field of view and a critical spatial sampling rate. Their primary design focus is on sky survey projects geared towards detecting near-Earth objects or bright supernovae. Due to the difference in transmission band design and aperture size, the optical telescope has a higher throughput compared to the near-ultraviolet telescope. Given that the dark current of CMOS cameras increases with temperature, temperature control is necessary to minimize noise and ensure observation quality. However, in a vacuum environment, natural heat dissipation is ineffective. Furthermore, due to limited space and power constraints on the satellite, there is no active cooling or shielding system in place. Consequently, heat can accumulate during camera operation, causing the temperature to rise from -20 to 30 degrees. To solve this problem, we set an upper exposure time limit and limit the camera to capture 100-120 frames after starting to work to maintain the temperature within an acceptable range.\\

\begin{table}[]
\caption{Parameters of visible and near-ultraviolet telescope onboard the Yangwang-1 satellite.}
\label{tab:parameter}
\begin{center}
{%
\centering
\begin{tabular}{|c|cc|}
\hline
Telescope                 & \multicolumn{1}{c|}{Optical}       & Near-ultraviolet     \\ \hline
Detector             & \multicolumn{2}{c|}{CMOS Gsense400BSI} \\ \hline
Optical System       & \multicolumn{2}{c|}{Refractive Telescope}                 \\ \hline
Observation Band     & \multicolumn{1}{c|}{420-700nm}     & 250-280nm       \\ \hline
Diameter       & \multicolumn{1}{c|}{75mm}          & 90mm            \\ \hline
Field Of View        & \multicolumn{1}{c|}{8.8°×8.8°}     & 5.7°×5.7°       \\ \hline
Focal Length         & \multicolumn{1}{c|}{146.6mm}       & 207.7mm         \\ \hline
Transmittance        & \multicolumn{1}{c|}{93.67\%}       & 11.5\%          \\ \hline
Single Exposure time & \multicolumn{2}{c|}{10ms-4800ms}                     \\ \hline
Distortion           & \multicolumn{2}{c|}{$\le 3\%$}                            \\ \hline
\end{tabular}}
\end{center}
\end{table}

Due to temperature fluctuations, the dark current of the camera undergoes significant changes. This issue is further exacerbated by the small aperture of telescopes, resulting in limited number of stars with a sufficient signal-to-noise ratio, particularly in the data obtained by the near-ultraviolet telescope. Both of these CMOS cameras utilized in the satellite consist of $2048\times 2048$ pixels with a quantum efficiency of approximately 70\% to 80\%. More detailed parameters can be found on the official site \footnote{\url{https://www.gpixel.com/en/pro_details_1196.html}}. Unfortunately, the camera chip on the satellite does not possess an attached thermometer. Although several thermometers are installed elsewhere, they fail to accurately reflect the temperature of the chip. Therefore, we propose obtaining the relationship between the dark current level and the temperature through laboratory data and developing a data-driven dark current noise model to estimate the dark current level. Further discussion regarding the model will be provided in Section~\ref{sec:CMOS}.\\

This paper utilizes two different types of data: laboratory test data and real observation data from the satellite. The laboratory test data are employed to construct the data-driven model. Due to the tight schedule involved in the satellite's manufacturing process, we have had only a few days available to acquire the test data. Consequently, we have established a chamber in the laboratory to capture dark current data for the camera. This chamber effectively blocks light and simulates the observation environment. Within the chamber, we have placed the CMOS camera and obtained dark frames. During observations, the CMOS camera continuously captures 100-120 images with an exposure time of 4 seconds for the near-ultraviolet telescope and an exposure time of 1 second for the optical telescope. Following this, the camera is turned off for a period to allow the temperature to decrease. To replicate the observation strategy, we have conducted the same procedure in the laboratory to obtain the dark current frames. Each iteration involved acquiring 100-120 frames consecutively, followed by turning off the camera to ensure temperature reduction. We have repeated this process for five rounds on each camera, resulting in a total of 5 rounds of testing with 100-120 frames in each round. It is important to note that, in this paper, the frame sequence number serves as a proxy for temperature since the temperature increases as soon as image capture begins. For other different applications, different definitions may be employed.\\ 

\begin{figure} [ht]
   \begin{center}
   \begin{tabular}{c} 
   \includegraphics[height=4cm]{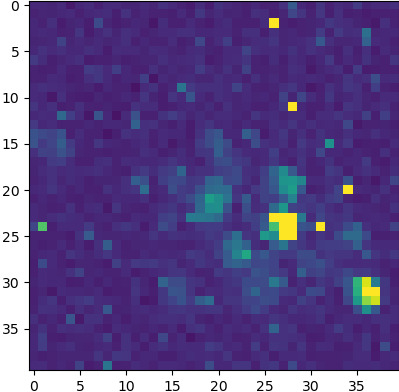}
   \includegraphics[height=4cm]{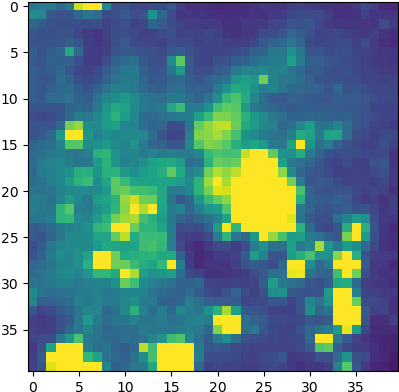}
   \end{tabular}
   \end{center}
    \caption
   { \label{fig:01}The same sky area captured by the near-ultraviolet telescope (the left panel) and the optical telescope (the right panel), when they are pointing to $RA=148.88822$ and $Dec=69.06529$. Since these two telescopes have different spatial resolutions, these two images have different field of views ($0.11^{\circ} \times 0.11^{\circ}$ for optical images and $0.17^{\circ} \times 0.17^{\circ}$ for near-ultraviolet images).}
 \end{figure}

The Yangwang-1 satellite has collected real observation data shortly after its launch, spanning from June 14, to June 28, 2021. In this paper, the data are obtained correspond to the telescope pointing at $RA=148.88822$ and $Dec=69.06529$. A total of 1347 images have been obtained and around 120 images were obtained each day. Figure~\ref{fig:01} shows both an optical image and a near-ultraviolet image of the same area of the sky. From the figure, it is evident that there are significantly fewer stars visible in the near-ultraviolet band compared to the optical band (white light). This disparity may be attributed to the narrow bandwidth of the near-ultraviolet filter. Additionally, the figure reveals that images captured by the near-ultraviolet telescope are critically sampled, and the quality of celestial object images is notably affected by dark-current noise and bad pixels.  \\

\section{Data-Driven Dark Current Noise Model of the CMOS Camera} \label{sec:CMOS}
\begin{figure} [ht]
   \begin{center}
   \begin{tabular}{c} 
   \includegraphics[height=12cm]{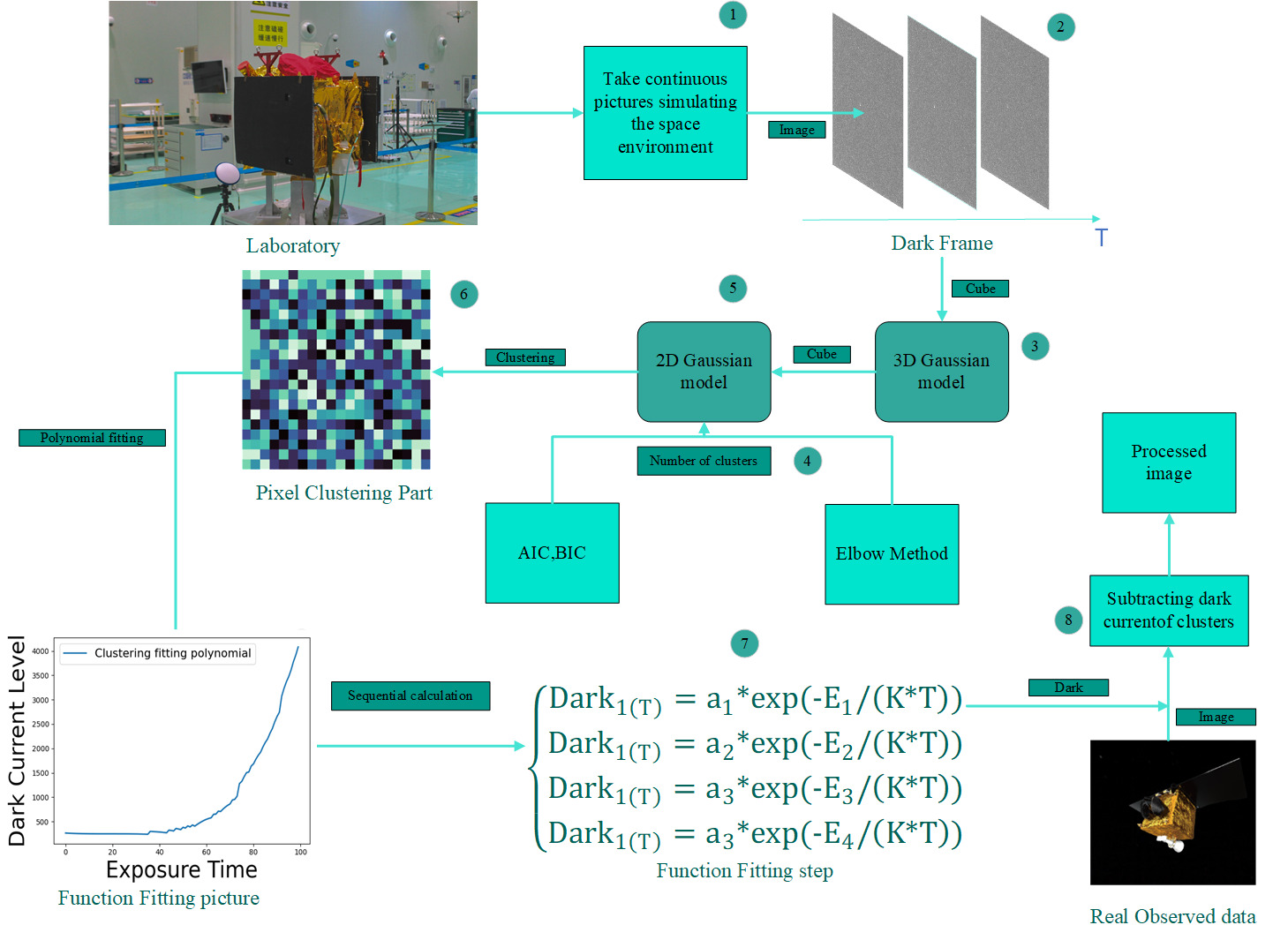}
   \end{tabular}
   \end{center}
    \caption
   {\label{fig:0} The schematic of the data-driven dark current noise model of the CMOS camera encompasses two essential steps: the pixel clustering step and the function fitting step. Initially, we collect test data following the observation strategy. Subsequently, we cluster the pixels based on their temperature (frame sequence number)-dark current relations. Using the clustering results, we fit functions to each cluster, representing the temperature (frame sequence number)-dark current relationship specific to the CMOS camera. In real observations, we start by acquiring the CMOS temperature (frame sequence number) using the gray scale values of non-celestial object pixels. This information enables us to determine the dark current for all pixels in the CMOS camera.}
 \end{figure}

The flowchart shown in Figure~\ref{fig:0} outlines the procedure for building and applying the data-driven dark current noise model of the CMOS camera. The flowchart consists of two primary steps: a pixel clustering step and a function fitting step. Within the pixel clustering step, pixels are clustered together based on their temperature-dark current relations. Following that, a function is fitted for each cluster to establish the temperature (frame sequence number)-dark current relation. The following subsections will provide a detailed discussing of these steps.\\

\subsection{Principle of the Data-Driven Dark Current Noise Model}
\label{sec:principle}
The data-driven dark current noise model is built considering the specific requirements of the application and the underlying physical processes. In the imaging process of the CMOS camera, photons are initially converted into electrons, which are then converted into digital numbers. However, due to the thermal movement of electrons, some electrons reach the electrodes randomly, resulting in dark current noise in the captured images, even if there are no photons \citep{magnan2003detection}. Theoretically, the dark current is directly related to the activation energy and the dark current level follows the Arrhenius law \citep{widenhorn2002temperature}. The dark current level is defined by factors such as temperature, exposure time, and the characteristics of the CMOS camera. As the exposure time or temperature increases, the dark current also tends to increase. However, since the dark current is generated randomly, the non-uniformity of the dark current manifests itself as a fixed mode noise with a random pattern specific to each CMOS camera, which can be observed in long-exposure images \citep{pain2005excess}.\\

Given that every pixel in the CMOS camera undergoes the same manufacturing process within a single chip, the dark current level in each pixel is expected to exhibit a similar trend with temperature, albeit with varying parameters. Based on this understanding, we put forward the following assumptions for the data-driven model of the CMOS camera:\\
When the CMOS camera reaches a specific temperature, its thermal distribution stabilizes.\\
The dark current of each pixel in the CMOS camera follows a random distribution.\\
The relationship between the average dark current level and temperature can be approximated using a set of functions.\\
The average dark current in each pixel will vary continuously as the temperature on the chips changes continuously.\\

Based on aforementioned assumptions, we can consider the dark current level at different temperatures in each pixel as one-dimensional random vectors, denoted as $X$ (given temperature). Furthermore, in the specific application we are addressing in this paper, we make the assumption that the temperature rises as the camera starts to capture images. We also apply a uniform exposure strategy across our observations, which is why we designate the frame sequence number, denoted as 'T,' to represent the temperature. However, in various other application scenarios, we can adjust our choice of parameters accordingly.\\

It may seem plausible to fit the relation between them using individual functions for each pixel. Nevertheless, the dark current levels exhibit random variations during each observation, and fitting a function for each pixel might result in overfitting problems. This could introduce additional errors in practical applications \citep{bashir2020information}. To address this issue, two potential solutions emerge: the first involves conducting multiple tests (with much large number than the number of free parameters in the function) for the same chip and collecting these data to derive the relationship by fitting gray scale values from all these measurements, while the second option entails collecting all pixels displaying similar relations between gray scale values and frame sequence number and then fitting the relationship between gray scale values and frame sequence number with measurements from all these pixels. Due to our limited time for laboratory testing, we opt for the second approach.\\

We propose clustering the pixels based on their frame sequence number - dark current relations and then build the relation between the dark current level and frame sequence number with data from these pixels. By clustering them according to the dark current and frame sequence number data, we can fit their relations using a smaller number of functions compared to the total number of pixels. This approach allows us to achieve effective representations based on statistical learning theory. Taking into account the aforementioned principle, we establish the framework discussed in the subsequent sections.\\

\subsection{The Pixel Clustering step}
We employ the Gaussian Mixture Model (GMM) to classify pixels into different clusters based on the relationship between dark current and temperature. The GMM is a widely used clustering method that can model various distributions, even when they deviate from a strict Gaussian distribution \citep{rasmussen1999infinite, xuan2001algorithms, Lee2012, Zhang2016, Jia2018}. In this study, we assume that the dark current vectors $X=\lbrace X_1,X_2,\cdots,X_M \rbrace$ are a linear combination of $N$ Gaussian functions. The latent variables are represented by a matrix of size $N \times M$. Consequently, the conditional probability of a dark current vector $x_i$ belonging to class $c_j$ follows a Gaussian distribution:
\begin{equation}
p(x_i=c_i|c_i)=G(x_i;\mu_i,\Sigma_i)=\frac{1}{(2\pi)^\frac{D}{2} |\Sigma_i|^\frac{1}{2}}exp(-\frac{(x_i-u_i)^T\sum^{-1}(x_i-u_i)}{2}),
\end{equation}
where $G(x_i;\mu_j,\sum _j)$ is a standard Gaussian distribution with $\mu$ as the mean value and $\Sigma$ as the covariance matrix. Thus, the probability density function of the Gaussian mixture model can be obtained by summing the combined distribution of all possible states:\\
\begin{equation}
\begin{split}
p(X|\lambda)=\sum_{i=1}^N p(c_i)\cdot p(x_i|c_j)=\sum_{i=1}^N Z_i G(x_i;\mu_i,\Sigma_i),\\
\lambda = \{Z_i,\mu_i,\Sigma_i\},
\end{split}
\end{equation}
where $Z_i$ is the weight of a Gaussian function in the mixed model and $\sum_{i=1}^N Z_i=1$. Given the GMM and the number of clusters $N$, we would fit parameters $\lambda$ in the GMM with the training data using the expectation maximization:\\
1. We initialize the parameters of $N$ Gaussian functions, including their means $\mu_i$, variances $\Sigma_i$, and weights $Z_{jc}$, with random numbers.\\
2. In the expectation step, we calculate the membership degrees of a dark current vector $x_i$ for each cluster $c$ using equation~\ref{eq3},
\begin{equation}
\label{eq3}
\begin{split}
    E[Z_{ic}]=\frac{p(x_i|\mu_c,\Sigma_c)}{\sum_{c=1} ^N p(x_i|\mu_c,\Sigma_c)},\\
    p(x_i|\mu_c,\Sigma_c)=\frac{1}{\sqrt{2\pi}\Sigma_c}exp(-\frac{(x_i-\mu_c)^2}{2\Sigma_{c}^2}),
\end{split}
\end{equation}
where $p(x_i|\mu_c,\Sigma_c)$ is probability of a pixel belonging to a cluster. $Z_{ic}$ is a belonging function, which is defined with equation~\ref{eq4}:
\begin{equation}
\label{eq4}
Z_{ic}=\left\{
\begin{array}{lr}
1, \text{ if the dark current - frame sequence number vector of a pixel i belongs to the Gaussian function $c$} & \\
0, \text{ else} &
\end{array}
\right.
\end{equation}
3. In the maximization step, we update the parameters of all $M$ Gaussian functions as well as the weights of each Gaussian function based on all the dark current-frame sequence number vectors using equation~\ref{eq5}:
\begin{equation}
\label{eq5}
\begin{split}
\mu_c = \frac{\sum_{i=1} ^m E[Z_{ic}*y_i]}{\sum_{i=1} ^m E[Z_{ic}]},\\
\Sigma_c = \frac{\sum_{i=1} ^m E[Z_{ic}](X-\mu_c)(X-\mu_c)^T}{\sum_{i=1}^m E[Z_{ic}]},\\
   W_c=\frac{\sum_{i=1} ^m E[Z_{ic}] }{m}.
\end{split}
\end{equation}
4. We will repeat steps 2 and 3 iteratively until the log-likelihood function defined in equation~\ref{eq6} reaches convergence, when $F$ does not decrease by 2 iterations.\\
\begin{equation}
\label{eq6}
F=\sum_{i=1}^N log( p(x_i|\mu_c,\Sigma_c)).
\end{equation}

\begin{figure} [ht]
   \begin{center}
   \begin{tabular}{c} 
   \includegraphics[height=5cm]{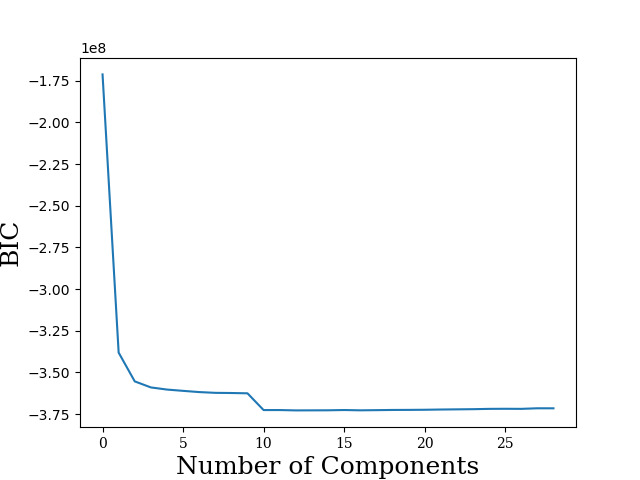}
   \includegraphics[height=5cm]{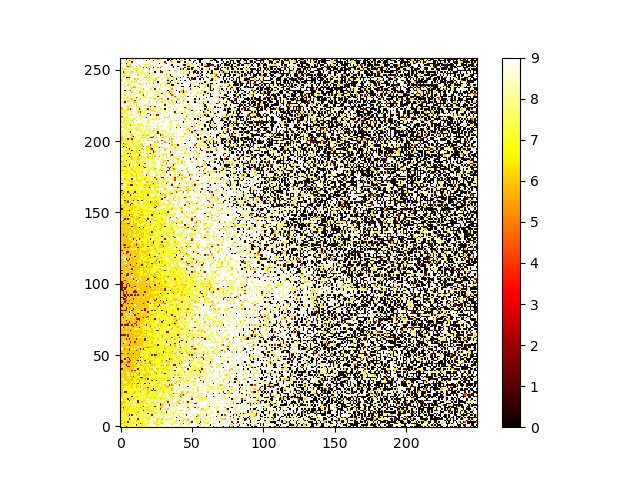}
   \end{tabular}
   \end{center}
    \caption
   { \label{fig:bic} The left panel displays the BIC values corresponding to various clusters. The right panel shows the distribution of pixels across different clusters.}
 \end{figure}

Using the aforementioned GMM model, we are able to identify clusters of pixels that exhibit the same relationship between dark current and frame sequence number. However, it is important to note that the number of clusters is a hyperparameter that requires careful selection. To determine the optimal number of clusters, we employ the Bayesian Information Criterion (BIC) as evaluation criteria \citep{burnham2004multimodel}. We utilize the elbow method, as suggested by \citet{thorndike1953belongs}, to find the optimal number of clusters. By incrementally increasing the number of clusters, we evaluate the BIC defined in equation \ref{eq7}:
\begin{equation}
\label{eq7}
\begin{split}
BIC=c*ln(Num)-2ln(L),
\end{split}
\end{equation}
where $c$ is the number of clusters, $L$ is corresponding likelihood function and $Num$ is the number of targets in a particular cluster. $L$ is defined in:
\begin{equation}
\label{eq8}
 L= -(Num/2)\ast \ln{(2\ast \pi)}-(Num/2)\ast ln{(sse/Num) }-Num/2,
\end{equation}
where $sse$ represents the discrepancy between the gray scale values and the prediction results obtained from the functions. As the number of clusters ($c$) increases, the number of samples assigned to each category decreases, resulting in a decrease in the BIC value. We determine the optimal number of clusters when the reduction in BIC decreases significantly, as noted in \citet{bholowalia2014ebk,syakur2018integration}. In practical applications, we explore different numbers of clusters for multiple clustering iterations to identify the optimal number of clusters. The BIC curve is shown in the left panel of Figure~\ref{fig:bic}. With the elbow method, we select 10 as the optimal number of clusters, because the BIC stops decreasing when the number of components is larger than 10. With the pixel clustering step, we can group pixels into distinct clusters, where all pixels in the same cluster share a common relationship between dark current and frame sequence number. The map illustrating the distribution of different clusters can be seen in the right panel of Figure~\ref{fig:bic}. As depicted in this figure, our approach effectively identifies areas with glowing characteristics and adjacent regions that display similar dark current properties.\\

\subsection{The Function Fitting Step} \label{sec:polfit}
While we have established the relationship between dark current and frame sequence number for each pixel during the pixel clustering step using the GMM model, it is essential to create a statistical model that effectively characterizes this relationship for pixels within the same cluster. Furthermore, this step allows us to pinpoint any defective or anomalous pixels. To tackle these challenges, we will employ the following procedures to perform function fitting and detect problematic pixels.\\
1. We determine the bias level by analyzing the image captured with the shortest exposure time of 10 ms, and all subsequent analyses are conducted using bias-subtracted images.\\
2. Mean values ($Mean_i$) could be calculated for all pixels belonging to the same cluster $i$, with one dimension representing $Mean_i$ and the other dimension representing the frame sequence number ($T$).\\
3. Following the Arrhenius law, we employ the $\frac{1}{T}$ as input and the natural logarithm of $Mean_i$ as the output for polynomials fitting. Different polynomials are fitted for pixels assigned to different clusters. \\
4. Iterations will be performed to fit these polynomials. In each iteration, pixels that exhibit significantly large deviations (5 times the mean square error) would be identified as bad or anomalous pixels. Subsequently, the polynomials would be fitted using the remaining pixels, and the mean square error could be calculated.\\
5. Bad pixels are promptly removed and excluded from subsequent image processing. We employ a mask matrix that designates all bad pixels as 0, while normal pixels assigned to various clusters will be tagged with their respective cluster indices.\\

\begin{figure} 
\begin{center}
   \includegraphics[height=5cm]{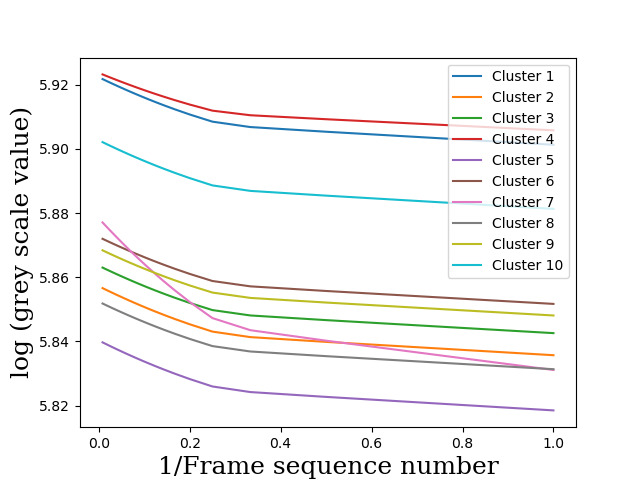}
   \includegraphics[height=5cm]{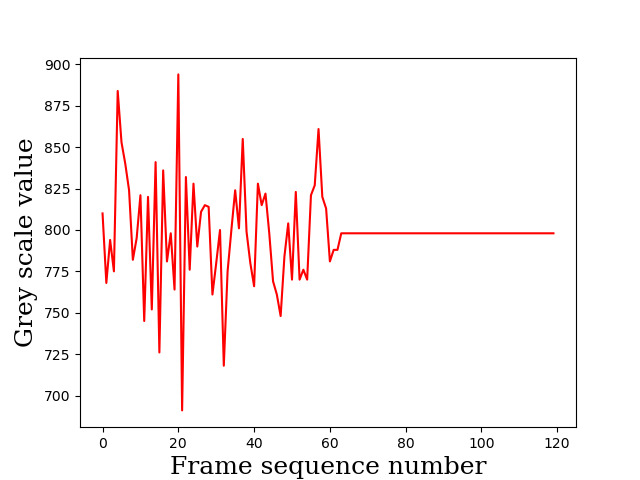}
\end{center}
    \caption{ \label{fig:1}The left panel shows the relation between logarithm of dark current level and 1/\ Frame Sequence Number. The right panel shows the relation between the dark current level and Frame Sequence Number.}
 \end{figure}

Following the steps outlined, we create an index matrix that assigns labels to pixels that share the same relation between dark current level and frame sequence number. Several functions representing these relationships are also generated. To visually depict the connection between the frame sequence number and the dark current level, we have plotted these relationships in the left panel of Figure~\ref{fig:1}. As depicted in this figure, all 10 clusters exhibit a consistent trend, highlighting the effectiveness of the selected polynomials. Additionally, we have identified a bad pixel and displayed its relationship with the dark current level and the frame sequence number in the right panel of Figure~\ref{fig:1}. As shown in this figure, the bad pixel exhibits random fluctuations until it reaches a frame sequence number of 63. Beyond this point, at sufficiently high temperatures, the pixel outputs a constant value of 800.\\

To evaluate the effectiveness of our approach, we analyze 100 consecutive frames of observation images captured by the near-ultraviolet telescope aboard the Yangwang-1 satellite. Our method, as described earlier, is utilized to determine the frame sequence number and the dark current level as the Fit method. For comparative analysis, we also employ two additional methods: 'Match,' representing the matching method, which selects the dark frame obtained on the ground close to observed images as the dark current level, and 'Per-fit,' representing the per-pixel fitting method, which matches each pixel with a distinct function constructed using ground test data, and the mean temperature (frame sequence number) is chosen as the index to derive the dark current level.\\

\begin{figure} [ht]
   \begin{center}
   \begin{tabular}{c} 
   \includegraphics[height=5cm]{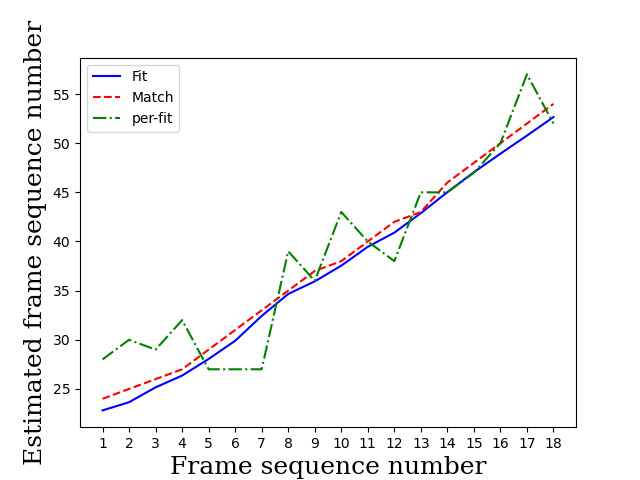}
   \includegraphics[height=5cm]{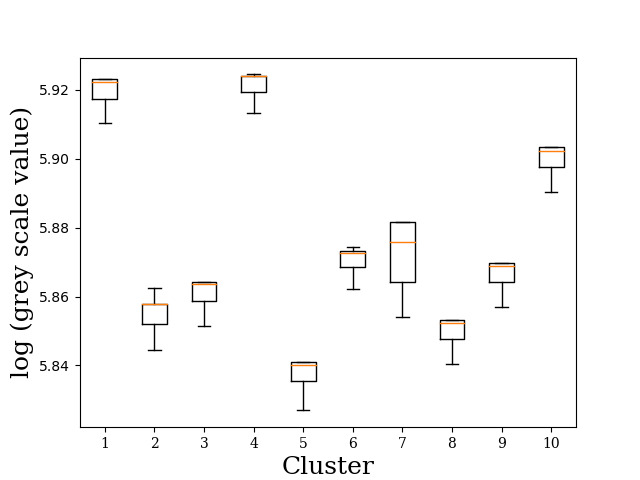}\\
   \centering
     \includegraphics[height=5cm]{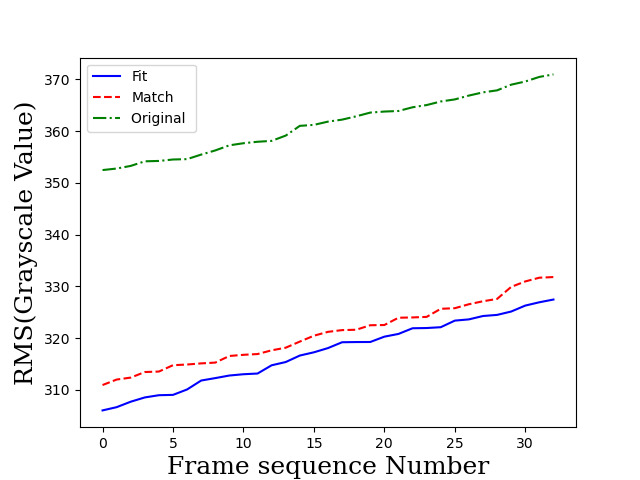}
   \end{tabular}
   \end{center}
    \caption
   { \label{figestimated} The top left panel shows the three methods('Fit', 'Match','Per fit') for predicting the true frame rate of the image. The top-right panel visualizes background levels of individual pixels within a single observation image for each cluster. The bottom panel shows the RMS comparison with the original image after processing by both methods 'Fit' and 'Match'.}
 \end{figure}

The results are illustrated in Figure~\ref{figestimated}. The top-left panel displays the estimated frame sequence number alongside the ground truth frame sequence number of real observation images. As depicted in this figure, both the Fit method and the Match method exhibit the same trend, wherein the estimated frame sequence number increases with the ground truth frame sequence number. Conversely, the Per-fit method fails to effectively capture the variations in the frame sequence number. Concurrently, the distribution of dark current levels is presented in the top-right panel of Figure~\ref{figestimated}. Notably, the difference between the dark current levels for the same cluster (obtained through the data-driven approach) is minimal, to the extent that they are scarcely discernible, even on a logarithmic scale. To highlight the distinctions between these methods, we have plotted the residual root mean square (RMS) of grayscale values in the bottom panel of Figure~\ref{figestimated}. As observed in this figure, both the Fit method and the Match method successfully reduce the residual RMS, and our Fit method reaches a lower RMS value.\\

In real applications, the data-driven approach includes both the index matrix and the functions that reflect the dark-current and temperature relations. When we use the data-driven approach to process real observation images, we can directly select a subset of pixels that do not capture photons from celestial objects and utilize their gray-scale values to estimate the camera's temperature. Then we can estimate the dark current of the whole detector. Further details on the application of these models will be discussed in Section~\ref{sec:realapp}.\\

\section{Application of the Data-Driven Dark Current Noise Model} \label{sec:realapp}
\subsection{Data Processing with the Data-Driven Dark Current Noise Model} \label{sec:dataproc}
\begin{figure} [ht]
   \begin{center}
   \begin{tabular}{c} 
   \includegraphics[height=4cm]{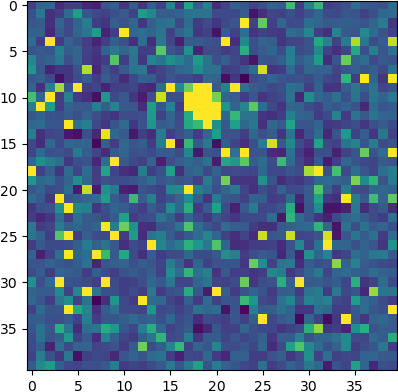}
   \includegraphics[height=4cm]{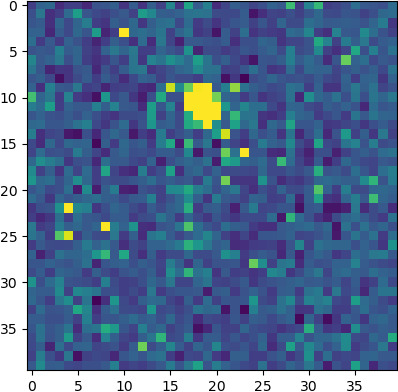}\\
    \includegraphics[height=4cm]{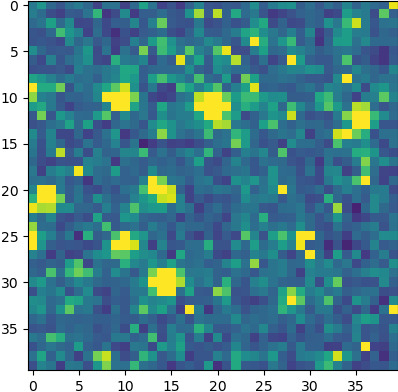}
    \includegraphics[height=4cm]{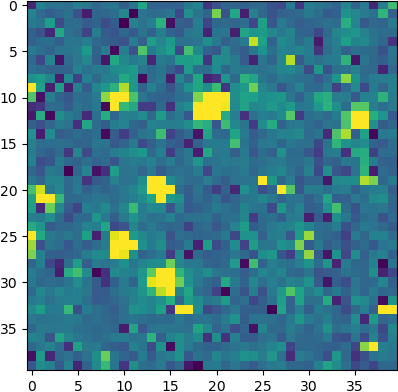}
   \end{tabular}
   \end{center}
    \caption
   { \label{fig:2} The left panel displays the original images and the right panel displays the processed images. The top panel shows images obtained by the near ultra-violet telescope, while the bottom panel shows images obtained by the optical telescope.}
 \end{figure}

For the YangWang Satellite, target detection constitutes the crucial first step in extracting scientific findings from observational data. However, the design of the satellite and observation mode result in dark current being the primary source of noise in both the near-ultraviolet and optical telescope images, as opposed to sky background or stray light. Therefore, accurately accounting for the dark current's contribution is essential. Following initial bias removal, we utilize the source detection algorithm SExtractor (with a connected area of 5 pixels and a 3 sigma detection threshold) to identify potential celestial object candidates. Subsequently, we employ a data-driven model to estimate the dark current using the remaining pixels. By utilizing around 10,000 pixels and applying the least squares method to fit the polynomials, we can determine the frame sequence number of the CMOS camera. Based on the frame sequence number, we could obtain the dark current frame for the CMOS camera with the data-driven model. Finally, through the subtraction of the dark current from the original image and masking of bad pixels (around 50000 pixels), we obtain processed images. Figure \ref{fig:2} displays the images captured by the near-ultraviolet telescope and the optical telescope, as well as the processed images. This figure demonstrate that our method successfully eliminates bad pixels and reduces the influence of dark current, indicating its potential to enhance the detection limit of the telescope.\\

With processed images, we analyze the detection efficiency and the detection accuracy with catalog from Gaia DR2 for the optical telescope, and with catalog from the GALEX for the near-ultraviolet telescope. Given the relatively small apertures of the telescopes used in this study, their detection capabilities are somewhat limited. We expect that all stars detected by these telescopes should also be within the Gaia catalog with apparent magnitudes below 19 and all the GALEX catalog. Therefore, we define the detection efficiency as the percentage of stars that our algorithm successfully identifies. Conversely, stars detected by our source detection algorithm but not listed in Gaia DR2 or the GALEX catalog are considered false detection results, which are used to calculate the accuracy rate. It is important to note that the definition employed here may lead to the omission of a few targets missed by the Gaia, however these instances are expected to be rare. It should be mentioned that the detection efficiency and the detection accuracy are differential quantity, which are defined for stars with specific brightness. Therefore, we evaluate these two values by stars with each apparent magnitude as a provisional measure in this paper. Considering the improved image quality, we can expect higher detection efficiency rates and higher accuracy rates in processed images using the same source detection algorithm. In the subsequent subsections, we will present additional evidence of the enhancements achieved by our method when employing the SExtractor and a Faster RCNN based method for the detection of celestial objects.\\

\subsection{Target Detection Results With The SExtractor} \label{sec:detect}
\begin{figure} [ht]
\begin{center}
\begin{tabular}{c} 
\includegraphics[height=4cm]{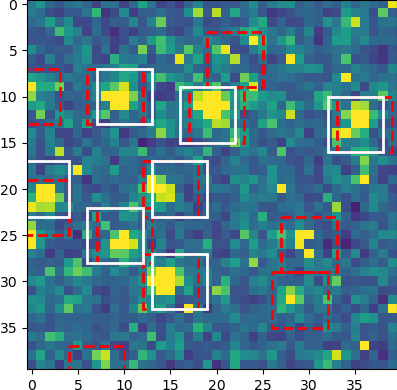}
\includegraphics[height=4cm]{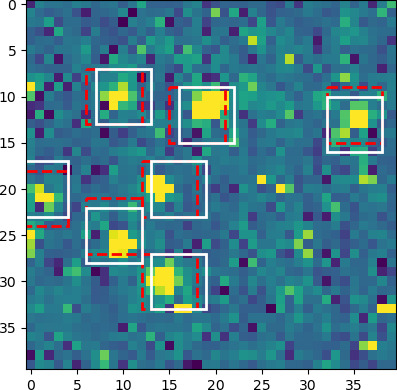}\\
\includegraphics[height=4cm]{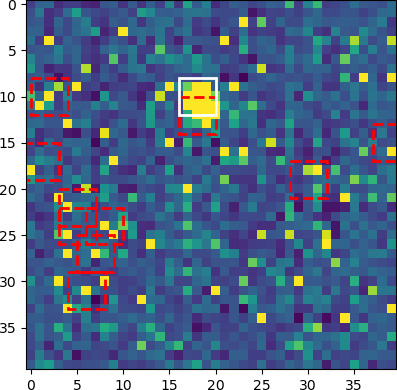}
\includegraphics[height=4cm]{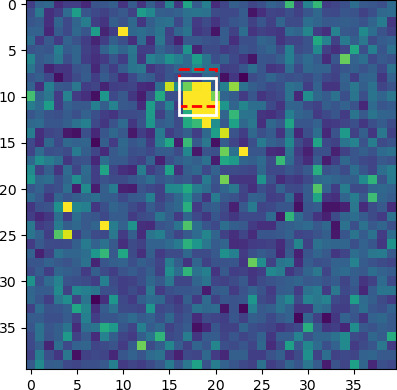}
\end{tabular}
\end{center}
 \caption
 { \label{fig:kj_sex} Part of images in optical (top panel) and near-ultraviolet band (bottom panel) before and after being processed with our method. White boxes with solid lines indicate positions of true celestial objects that are above the detection limit. Red boxes with dashed lines indicate positions of candidates detected by the SExtractor. As shown in this figure, the SExtractor could better detect celestial objects after being processed by our method (with much lower false detection rate).}
 \end{figure}

In this subsection, we employ the SExtractor as the source detection algorithm to identify celestial objects in both the original and images processed with our approach. The SExtractor is a widely recognized algorithm for source detection, well known for its reliable performance with various observational data sources \citep{bertin1996sextractor}. We use the same parameters for the SExtractor to process the original and processed images. Our method is applied to both the original near-ultraviolet and optical images. A portion of the original and processed images is displayed in Figure~\ref{fig:kj_sex}. In the figure, the white boxes represent detected targets, while the red boxes indicate true targets. As depicted, the SExtractor effectively identifies bright targets in both sets of images, but the presence of noise introduces some false detection results. After we process these images, we could obtain detection results with much lower false detection rates.\\

To further evaluate the performance of our algorithm, we conduct tests using an image obtained by the optical telescope. We use the SExtractor to detect targets from both the original and processed versions of the image. Subsequently, we set the limiting magnitude as $m=19$ and perform a cross-match with the Gaia DR2 catalog to analyze the detected results, as shown in the top panel of Figure~\ref{fig:pr}. The figure demonstrates significant improvements in both detection accuracy and efficiency achieved by the 'Fit' and 'Match' methods compared to baseline approaches. Notably, our 'Fit' method exhibits the largest increase in detection accuracy for stars across a broad range of brightness levels, encompassing both bright and dim stars.\\

\subsection{Target Detection Results With The Faster RCNN Based Source Detection Algorithm} \label{sec:fa}
\begin{figure} [ht]
   \begin{center}
   \begin{tabular}{c} 
   \includegraphics[height=5cm]{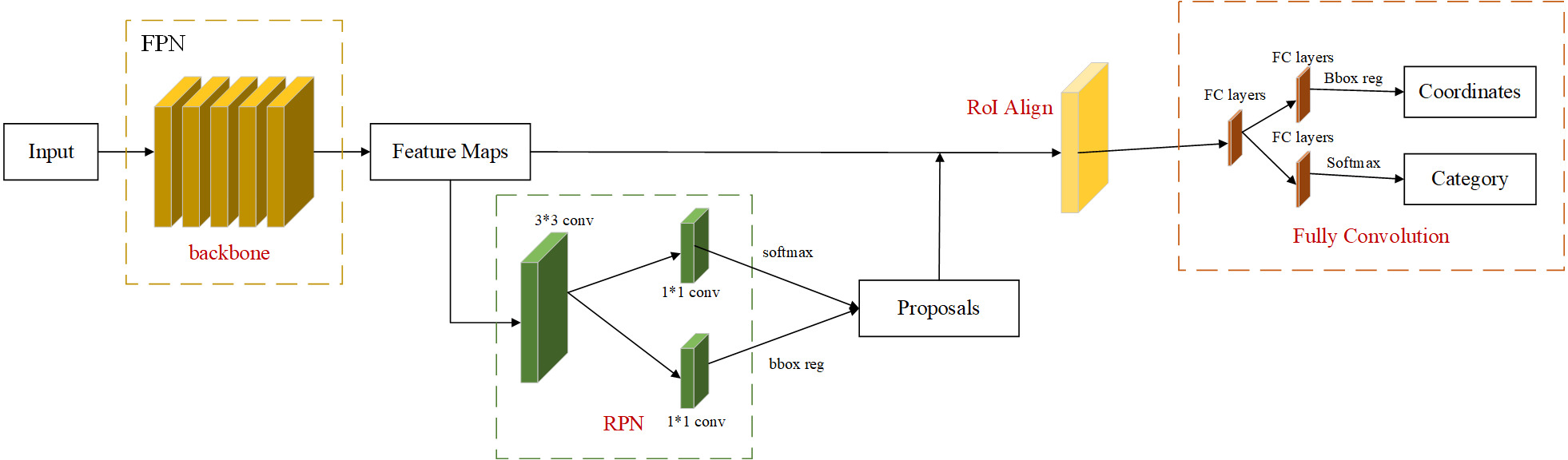}
   \end{tabular}
   \end{center}
    \caption
   { \label{figFasterrcnn} The structure of the Faster RCNN detection algorithm used in this paper.}
 \end{figure}

The Faster RCNN based celestial object detection algorithm, introduced by \citet{jia2020detection}, is designed specifically for detecting small and sparsely distributed celestial objects. The algorithm demonstrates superior detection capability for images of a specific type when provided with sufficient training data. For a comprehensive understanding of the method, see \citet{jia2020detection} and \citet{sun2023pnet}. Here, we provide a brief overview of the approach. As illustrated in Figure~\ref{figFasterrcnn}, the algorithm has the following steps:\\
Feature extraction: We employ a convolutional neural network (CNN) to construct a Feature Pyramid Network (FPN) that extracts features from various scales, facilitating subsequent detection. These feature maps are shared by both the Region Proposal Network (RPN) and the classification and regression neural network.\\
Region Proposal Networks (RPN): RPNs generate region proposals by processing feature maps. They identify regions in the images that are likely to belong to the foreground or background.\\
ROI Pooling: ROI Pooling can be applied to resize the proposed region to a normalized size, and then connected to a fully connected layer for prediction and regression.\\

 \begin{figure} [ht]
   \begin{center}
   \begin{tabular}{c} 
   \includegraphics[height=5cm]{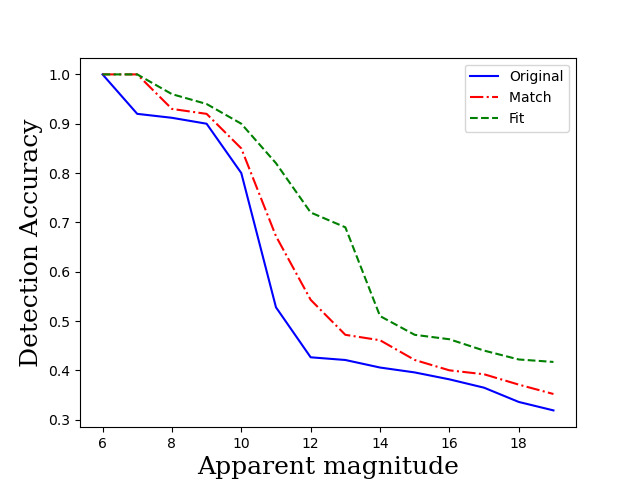}
   \includegraphics[height=5cm]{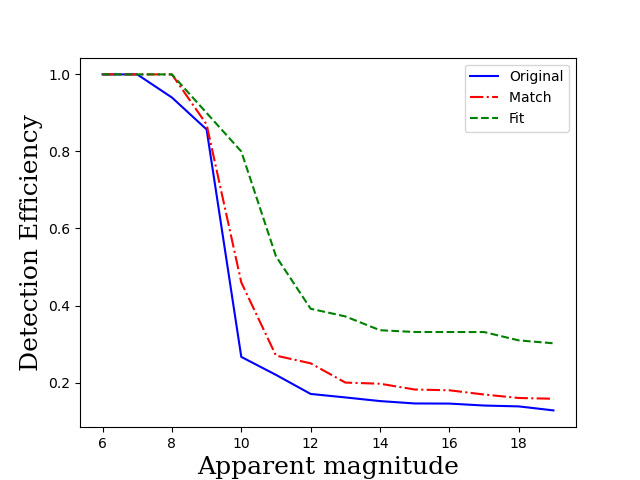}\\
   \includegraphics[height=5cm]{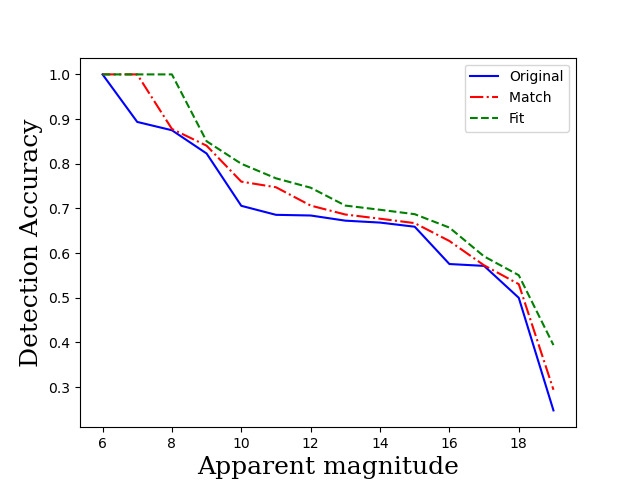}
   \includegraphics[height=5cm]{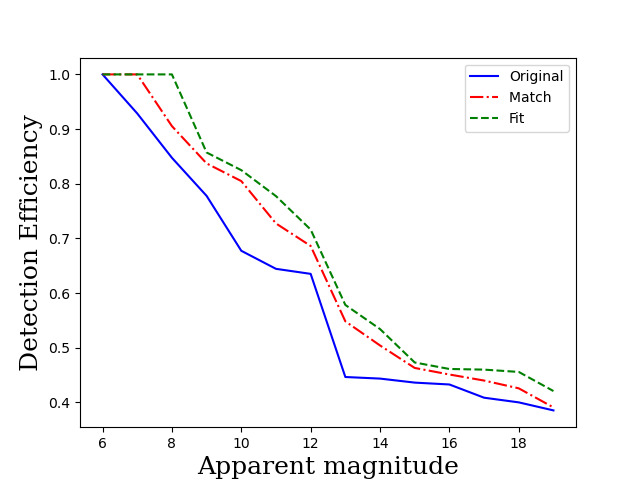}\\
    \includegraphics[height=5cm]{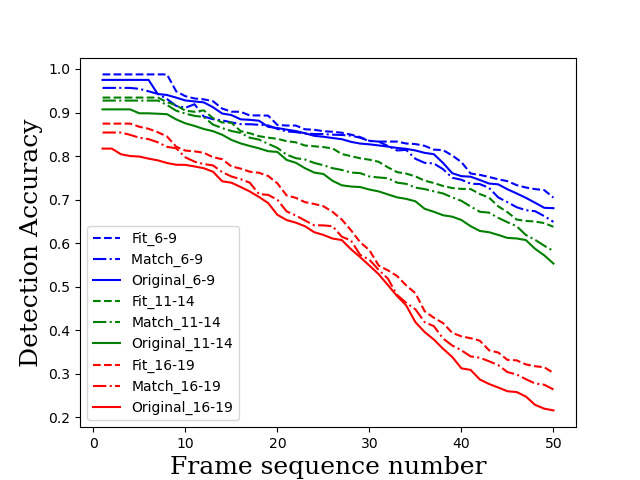}
   \includegraphics[height=5cm]{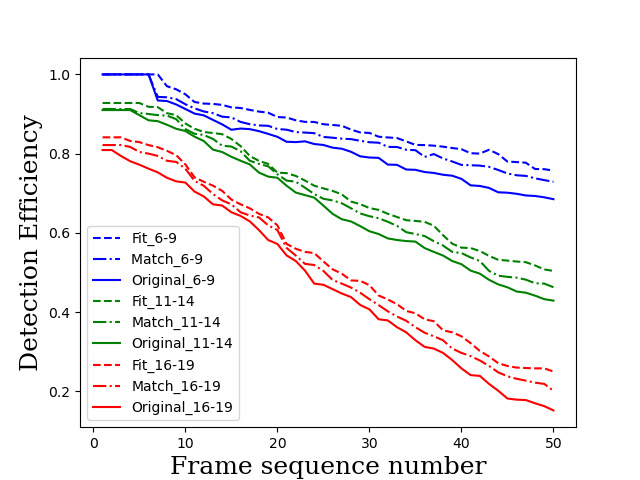}
   \end{tabular}
   \end{center}
    \caption
   { \label{fig:pr}The top and middle panels illustrate the detection accuracy and efficiency of both the SExtractor and Faster RCNN for stars with varying apparent magnitudes in the 47th. frame of images. The bottom panel shows how the detection accuracy and efficiency decrease with increasing temperature (frame sequence number).}
 \end{figure} 

We have trained two Faster RCNN models separately using original images and processed images from the optical telescope. A total of 200 frames of each category have been used as training data and positions of stars in these images are obtained by the Gaia DR2 catalog as labels. The Faster RCNN models have been trained for 30 epochs using the Adam optimizer with an initial learning rate of 0.001. Training the neural networks has taken 2-3 hours in a computer with one RTX 3090 GPU. Once trained, we have applied these neural networks to process images in the test dataset consisting of 50 frames of continuously observed images obtained by the optical telescope. For comparison, we have plotted the detection accuracy and detection efficiency for stars with different apparent magnitude in the middle panel of Figure~\ref{fig:pr}.\\

In the top panel of Figure~\ref{fig:pr}, significant performance enhancements are observed when employing the method to reduce the dark current noise, when the SExtractor is used as the detection algorithm. The same results could be observed in the middle panel of Figure~\ref{fig:pr}, when we use the Faster RCNN as the detection algorithm. For both the classical and the deep learning based source detection algorithm, the detection capability has notably increased by a minimum of 50\% for stars with a apparent magnitude exceeding 12, at a cost of a slightly reduce in detection accuracy. As discussed earlier, the rise in camera temperature during observations elevates the significance of dark current levels, particularly in images captured at higher frame sequence numbers. The bottom panel of the Figure~\ref{fig:pr} displays the detection accuracy and efficiency for images with varying frame sequence numbers. As depicted, the detection accuracy experiences a rapid decline for frame numbers exceeding 30, while the detection efficiency decreases steadily across different frame sequence numbers. As evident, dim stars suffer more significantly from the dark current issue, leading to reduced detection accuracy and efficiency. Nevertheless, our method effectively mitigates this problem.\\

\section{CONCLUSIONS AND FUTURE WORK} \label{sec:con}
In recent years, there has been a growing interest in launching small satellites for astronomical observations. To optimize costs, many of these satellites are opting for commercial cameras without dedicated cooling or shielding systems due to space and power limitations. In addition, these satellites aim to minimise additional observation and data transfer requirements. Addressing these needs, we propose an innovative approach to construct a data driven model of the CMOS camera, mitigating the effects of dark current and bad pixels. This method only requires some additional ground-based tests and data modeling processes. We validate our method using real observation data obtained from the near-ultraviolet and optical telescope onboard the Yangwang-1 satellite. The results demonstrate that our approach effectively reduces dark current noise and identifies bad pixels. Moreover, when combined with classical and deep neural network based target detection algorithms, our method enhances detection efficiency. This indicates that our approach can serve as an effective tool for CMOS cameras installed in future space-based satellites.\\

There are several aspects that require further investigation for our method. First, due to our tight time schedule, the data acquisition process is designed based only on experience. To enhance our approach, it would be beneficial to construct a finite element model of the camera and the telescope and develop an agent that can autonomously determine the optimal data acquisition strategy. Second, as the Yangwang-1 primarily experiences a dark current as the dominant noise source, we do not address the sky background or stray light in this paper, which could potentially impact the photometry accuracy. In the future, we plan to develop new techniques to separate and account for different sources of noise, while also expanding the noise modeling procedure for other space telescopes, such as cameras in the China Space Station Telescope (CSST) \footnote{\url{https://csst-tb.bao.ac.cn/code/csst_sim/csst-simulation}}.\\

\section*{Acknowledgements}
Authors would like to thank the reviewer for his/her kindly suggestions, which greatly improve the quality of this paper. The code used in this article can shared in the PaperData repository powered by China-VO with DOI of 10.12149/101386. This work is supported by the National Natural Science Foundation of China (NSFC) with funding numbers 12173027, 12303105 and 12173062, National Key Research and Development Program with funding numbers 2023YFF0725300. We acknowledge the science research grants from the China Manned Space Project with NO. CMS-CSST-2021-A01.
\bibliographystyle{aasjournal} 
\bibliography{sample631} 

\end{document}